\documentclass[
    ,final            
  ]
  {aipproc}

\layoutstyle{8x11double}

\newcommand{\lsim}{\raise.35ex\hbox{$<$}\kern-0.75em\lower.5ex\hbox{$\sim$}}
\newcommand{\gsim}{\raise.35ex\hbox{$>$}\kern-0.75em\lower.5ex\hbox{$\sim$}}

\newcommand{\simle}
{\raisebox{-0.75ex}[-1.5ex]{$\;\stackrel{<}{\sim}\;$}}

\def\d{{\partial}}
\def\s{{\sigma}}
\def\e{{\epsilon}}
\def\k{{ {\bf k} }}
\def\p{{ {\bf p} }}

\def\w{{\omega}}

\begin{document}

\title{Physical Meaning of the Current Vertex Corrections:
DC and AC Transport Phenomena in High-$T_{\rm c}$ Superconductors
}


\author{Hiroshi Kontani}{
  address={Department of Physics, Nagoya University,
 Nagoya, 464-8602, Japan}
}

\begin{abstract}
Famous non-Fermi liquid-like behaviors of
the transport phenomena in high-$T_{\rm c}$ cuprates
($R_{\rm H}$, $\Delta\rho/\rho$, $S$, $\nu$, etc)
are caused by the current vertex corrections in neary antiferromagnetic (AF)
Fermi liquid, which was called the backflow by Landau.
We present a simple explanation why the backflow is prominent
in strongly correlated systems.
In nearly AF Fermi liquid,
$R_{\rm H}$ is enhanced by the backflow because it 
changes the effective curvature of the Fermi surfaces.
Therefore, the relaxation time approximation is not
applicalbe to a system nearby a magnetic quantum critical point (QCP).

\end{abstract}

\maketitle


In the present article, we explain that the backflow
always plays important roles in Fermi liquids.
In a Fermi liquid,
excited quasiparticles (QP) interact with each other.
Landau showed that the QP energy at $\p$
in the presence of QP excitations is expresed as
 \cite{Landau}
\begin{eqnarray}
{\tilde \e}_\p= \e_\p+\sum_\k f_{\p,\k}\delta n_{\k} ,
\label{eqn:Landau}
\end{eqnarray}
where $\e_\p=\p^2/2m^\ast$ and $\delta n_{\k}$
represents the deviation of the QP distribution 
function from the ground state.
$f_{\p,\k}$ is the QP interaction (i.e., Landau interaction function)
which arises from the electron-electron correlations.
Hereafter, we consider the paramagnetic state.

\begin{figure}
  \includegraphics[height=.15\textheight]{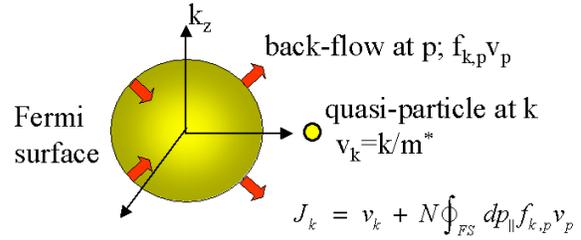}
  \caption{Backflow is the induced QP velocity over the FS 
 due to other QP excitations, which is one of the most
 essential properties of Fermi liquids.}
\label{fig:FS}
\end{figure}

First, we discuss an isotropic Fermi liquid.  
Here, we consider to add a QP just above the Fermi surface (FS) 
at $\k$ as shown in Fig.\ref{fig:FS}, 
whose lifetime $\tau$ is infinitesimally long.
Then, the QP energy at $\p$ shifts by $f_{\p,\k}$,
which induces the change of QP velocity \cite{Landau}.
Its summation over the FS is given by
$\sum_{|\p|<k_{\rm F}} {\bf \nabla}_\p f_{\p,\k} =
 N\oint_{\rm FS} dp_{\parallel} f_{\k,\p}{\bf v}_\p$,
which is called the backflow by Landau.
In a spherical system, the QP velocity at $\k$ and the backflow
are given by ${\bf v}_\k=\k/m^\ast$ and $\frac13 F_1 \k/m^\ast$
($F_1$ being a Landau parameter), respectively.
Because $1+\frac13 F_1=m^\ast/m$ due to Galilei invariance,
the total current ${\bf J}_\k$ is $\k/m$
 \cite{Landau}.
Thus, the backflow dominates the QP velocity in a strongly
correlated Fermi liquid where $m^\ast \gg m$.

The backflow plays an important role in DC transport phenomena.
To show this fact, we study the current vertex correction 
in the hydrodynamic limit ($\w\tau \ll 1$) based on the microscopic 
Fermi liquid theory.
The resistivity of interacting electrons without umklapp scatterings
should be completely zero, as consequence of the momentum conservation law.
However, the relaxation time approximation (RTA)
gives a finite resistivity because 
$\rho^{\rm RTA}\propto \tau >0$.
Yamada and Yosida solved this discrepancy
by taking the backflow (i.e., the current vertex correction) 
into account; they succeeded in reproducing $\rho=0$ 
in the absence of the umklapp scatterings,
even when $\tau$ is finite
 \cite{Review}.

\begin{figure}
  \includegraphics[height=.27\textheight]{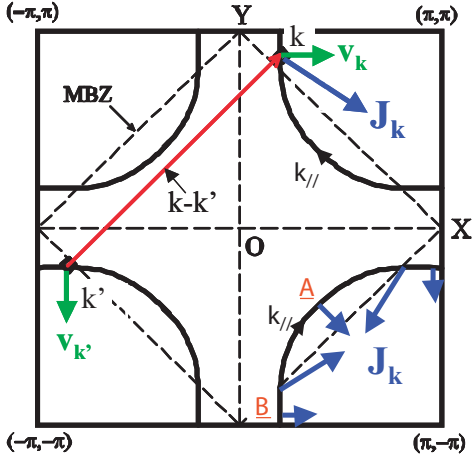}
  \caption{Schematic picture of ${\bf J}_\k$ in high-$T_{\rm c}$ 
 cuprates.
Because the excited QP at $\k$ reduces the QP energy $\e_{\k'}$
only for $\k'\approx \k+{\bf Q}$,
the induced beckflow is proportional to ${\bf v}_{\k+{\bf Q}}$.
 }
\label{fig:HTSC}
\end{figure}

In 1999, we found that the backflow in nearly 
antiferromagnetic (AF) metals plays highly nontrivial roles;
the total current ${\bf J}_\k$ is no more parallel to
the QP velocity ${\bf v}_\k$ due to the backflow
 \cite{Review,Kontani-Hall}.
The Bethe-Salpeter equation in the microscopic Fermi liquid theory 
is given by
\begin{eqnarray}
{\vec J}_\k &=& {\vec v}_\k + \oint_{\rm FS} dk_\parallel 
 \Gamma(\k,\p){\vec J}_{\p} ,
 \label{eqn:BS1}
\end{eqnarray}
where $\Gamma(\k,\p)$ corresponds to the Landau interaction
function in the hydrodynamic limit, and 
$k_\parallel$ is the momentum along the FS.
In the FLEX approximation,
$\Gamma(\k,\p) = \int d\e [{\rm cth}(\e/2T)-{\rm th}(\e/2T)]
{\rm Im}V^{\rm FLEX}_{\k-\p}(\e)\tau_\p$ and
$V^{\rm FLEX}_{\k}(\e) \propto \chi_\k^s(\e) \propto
\xi^2(1+\xi^2(\k-{\bf Q})^2 -i\e/\w_{\rm sf})^{-1}$,
where $\chi_\k^s$ is the spin susceptibility;
$\xi$ is the AF correlation length and ${\bf Q}=(\pm\pi,\pm\pi)$.
In the SCR or FLEX approximation,
$\xi^2\propto T^{-1}$ and $\xi\gg 1$ at lower temperatures.
Then, $\Gamma(\k,\p)$ takes a large value
only when $|\k-\p-{\bf Q}| \simle \xi^{-2}$.
As shown in Fig.\ref{fig:HTSC},
a QP excitation at $\k$ causes the backflow 
through the QP interaction $\Gamma$ in eq. (\ref{eqn:BS1}), 
only on the small portion of the FS around $\k'\approx \k+{\bf Q}$.
Then, the total current becomes
${\bf J}_\k \propto {\bf v}_\k + {\bf v}_{\k'}$,
so ${\bf J}_\k$ is not perpendicular to the FS except the points A and B,
which are on the symmetric axes.
In summary, the highly anisotropic Landau interaction
function $\Gamma(\k,\k')$ due to the AF fluctuations
causes the nontrivial $\k$-dependence of ${\bf J}_\k$.

\begin{figure}
  \includegraphics[height=.28\textheight]{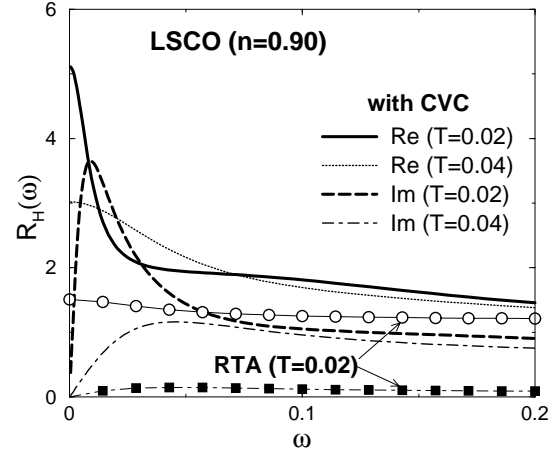}
  \caption{AC Hall coefficient obtained by the FLEX approximation.
$R_{\rm H}(\w)\approx 1/ne$ within the RTA,
whereas $R_{\rm H}(\w)$ by the conservation approximation
shows a prominent $\w$-dependence, which is consistent 
with experiments.
 }
\label{fig:IR-Hall}
\end{figure}

We show that the nontrivial behavior of ${\bf J}_\k$ 
shown in Fig. \ref{fig:HTSC} 
leads to the enhancement of $R_{\rm H}$.
The expression for the Hall conductivity is given by 
 \cite{Review}
\begin{eqnarray}
\s_{xy}^{\rm RTA}/B_z&=& |e|^3 \oint_{\rm FS}
 dk_\parallel |{\vec v}_\k \tau_\k|^2 C_\k^v,
\label{eqn:sxy-RTA} \\
\s_{xy}/B_z&=& |e|^3 \oint_{\rm FS}
 dk_\parallel |{\vec J}_\k \tau_\k|^2 C_\k^J , 
 \label{eqn:sxy}
\end{eqnarray}
where $k_\parallel$ is the momentum parallel to the FS,
$C_\k^v= -\d\theta_\k^v/\d k_\parallel$,
$C_\k^J= -\d\theta_\k^J/\d k_\parallel$,
$\theta_\k^{v}= \tan^{-1}(v_{\k x}/v_{\k y})$,
$\theta_\k^{J}= \tan^{-1}(J_{\k x}/J_{\k y})$, and
$\tau_\k= 1/2{\rm Im}\Sigma_\k(-i\delta)$ is the lifetime of QP. 
Because $C_\k^v$ represents the FS curvature (FSC) at $\k$,
we recognize the well-known result of the RTA;
{\it $\s_{xy}$ is determined by the FSC}.
However, this statement does not hold in nearly AF systems
any more, because the "effective FSC" in eq.(\ref{eqn:sxy}), 
$C_\k^J$, strongly deviates from the true FSC.
In Fig. \ref{fig:HTSC},
$C_\k^J>0$ around A whereas it is negative around B.
In contrast, $C_\k^v>0$ everywhere.
Moreover, the "cold spot" where $\tau_\k$ takes the maximum value on the FS
is around A (B) in the hole-doped (electron-doped) systems.
As a result, $\s_{xy}$ is positive (negative) in 
hole-doped (electron-doped) systems, as recognized in eq.(\ref{eqn:sxy}).
In fact, $R_{\rm H}$ in electron-doped systems is
negative although its FSC observed by ARPES is positive.
This fact cannot be explained by the idea, 
"deformation of the FS due to SDW transition",
because $R_{\rm H}<0$ even above $T_{\rm N}$.
In Ref. \cite{Kontani-Hall},
we solved this discrepancy by taking the fact that
the effective FSC around the cold-spot becomes negative
due to the backflow.
We have shown that $C_\k^J \propto \xi^2$
in both hole-doped and electron-doped systems
in terms of the conserving approximation,
so $R_{\rm H}$ in under-doped system is strongly enhanced 
at lower temperatures in proportion to $T^{-1}$.
Moreover, characteristic frequency and temperature dependences of 
the AC Hall coefficient
are also reproduced very well by taking the backflow into account,
as shown in Fig.\ref{fig:IR-Hall} \cite{Kontani-IR-Hall}.

In summary, the backflow is a natural consequence of the basic Fermi liquid equation, eq.(\ref{eqn:Landau}),
and it cannot be ignored in strongly correlated systems.
The total current in nearly AF metals 
is no more perpendicular to the FS due to the backflow,
because the Landau interaction function caused by the 
AF fluctuations has a prominent momentum dependence.
This fact means that the RTA is totally broken down.
The Hall coefficient shows a strong
temperature dependence because the backflow changes
the ``effective FSC'' around the cold spot.
By taking the backflow into account,
we succeeded in reproducing characteristic
non-Fermi liquid-like behaviors in various DC and AC transport phenomena
in high-$T_{\rm c}$ cuprates, even in the pseudo-gap region
 \cite{Review,Kontani-Hall,Kontani-Nernst}.

\end{document}